\documentclass[article,11pt]{revtex4}
\usepackage{graphicx} % standard LaTeX graphics tool;
\usepackage{amsmath} % For getting proper math eqns
\usepackage{amssymb} % Used for getting various symbols eg. gtrsim, lesssim etc.
\usepackage{bm} % For bold math (esp of lower greek letters)
\usepackage{dcolumn}% Align table columns on decimal point
\usepackage{color}
\usepackage{mathrsfs}
\usepackage{amsfonts}
\usepackage{varioref}
\usepackage{mathrsfs}
\usepackage{graphicx}
\usepackage{latexsym}
\usepackage{amsmath}
\usepackage{amssymb}
\usepackage{textcomp}
\usepackage{amsbsy}
\usepackage{graphics}
\usepackage{epstopdf}
\usepackage{color}
\usepackage[caption=false]{subfig}

\RequirePackage[colorlinks,citecolor=blue,urlcolor=magenta,linkcolor=blue]{hyperref}
\input epsf

\allowdisplaybreaks[4]
\begin{document}

\tolerance=5000

\title{Modified gravity as entropic cosmology}

\author{Shin'ichi~Nojiri$^{1,2}$\,\thanks{nojiri@gravity.phys.nagoya-u.ac.jp},
Sergei~D.~Odintsov$^{3,4}$\,\thanks{odintsov@ieec.uab.es},
Tanmoy~Paul$^{5}$\,\thanks{tanmoy.paul@visva-bharati.ac.in},
Soumitra~SenGupta$^{6}$\,\thanks{tpssg@iacs.res.in}} \affiliation{
$^{1)}$ KEK Theory Center, Institute of Particle and Nuclear Studies,
Oho 1-1, Tsukuba, Ibaraki 305-0801, Japan \\
$^{2)}$ Kobayashi-Maskawa Institute for the Origin of Particles
and the Universe, Nagoya University, Nagoya 464-8602, Japan \\
$^{3)}$ ICREA, Passeig Luis Companys, 23, 08010 Barcelona, Spain\\
$^{4)}$ Institute of Space Sciences (ICE, CSIC) C. Can Magrans s/n, 08193 Barcelona, Spain\\
$^{5)}$ Department of Physics, Visva-Bharati University, Santiniketan-731235, India\\
$^{6)}$ School of Physical Sciences, Indian Association for the Cultivation of Science, Kolkata-700032, India}

%\date{}

\tolerance=5000

\begin{abstract}
The present work reveals a direct correspondence between modified theories of gravity (cosmology) and entropic cosmology based on the thermodynamics of apparent horizon. It turns out that due to the total differentiable property of entropy, the usual thermodynamic law (used for Einstein gravity) needs to be generalized for modified gravity theories having more than one thermodynamic degrees of freedom (d.o.f.). For the modified theories having $n$ number of thermodynamic d.o.f., the corresponding horizon entropy is given by: $S_\mathrm{h} \sim S_\mathrm{BH} +$ terms containing the time derivatives of $S_\mathrm{BH}$ up to $(n-1)$-th order, and moreover, the coefficient(s) of the derivative term(s) are proportional to the modification parameter of the gravity theory (compared to the Einstein gravity; $S_\mathrm{BH}$ is the Bekenstein-Hawking entropy). By identifying the independent thermodynamic variables from the first law of thermodynamics, we show that the equivalent thermodynamic description of modified gravity naturally allows the time derivative of the Bekenstein-Hawking entropy in the horizon entropy.
\end{abstract}
%%%%%%%%%%%%%%%%%%%%%%%%%%%%%%%%%%%%%%%%%%%%%%%%%%%%%%%%%%%%%%%%%%%%%%%%%%%%%%%%%%%%%%%%%%%%%%%%%%%
%%%%%%%%%%%%%%%%%%%%%%%%%%%%%%%%%%%%%%%%%%%%%%%%%%%%%%%%%%%%%%%%%%%%%%%%%%%%%%%%%%%%%%%%%%%%%%%%%%%
%%%%%%%%%%%%%%%%%%%%%%%%%%%%%%%%%%%%%%%%%%%%%%%%%%%%%%%%%%%%%%%%%%%%%%%%%%%%%%%%%%%%%%%%%%%%%%%%%%%
%\newpage
%%%%%%%%%%%%%%%%%%%%%%%%%%%%%%%%%%%%%%%%%%%%%%%%%%%%%%%%%%%%%%%%%%%%%%%%%%%%%%%%%%%%%%%%%%%%%%%%%%%
%%%%%%%%%%%%%%%%%%%%%%%%%%%%%%%%%%%%%%%%%%%%%%%%%%%%%%%%%%%%%%%%%%%%%%%%%%%%%%%%%%%%%%%%%%%%%%%%%%%
%%%%%%%%%%%%%%%%%%%%%%%%%%%%%%%%%%%%%%%%%%%%%%%%%%%%%%%%%%%%%%%%%%%%%%%%%%%%%%%%%%%%%%%%%%%%%%%%%%%
%\pacs{}

\maketitle

\section{Introduction}\label{SecI}
The discovery of black hole thermodynamics associated with event horizon of the black hole puts two apparently different arenas of Physics, namely gravity and thermodynamics, on same footing \cite{PhysRevD.7.2333,Hawking:1975vcx,Bardeen:1973gs,Wald_2001}. In this regard, the entropy of black hole is generally considered to be the Bekenstein-Hawking type entropy (or have some other form), and the temperature of the black hole is given by the surface gravity of the same. Then the thermodynamic law of the black hole results to the underlying gravitational field equations.

In the context of cosmology, the Friedmann-Lema\^{i}tre-Robertson-Walker (FLRW) spacetime acquires an apparent horizon, and similar to black hole thermodynamics, the cosmological field equations are expected to derive from the thermodynamic laws of the apparent horizon \cite{PhysRevD.15.2738,Jacobson:1995ab,Cai:2005ra,Akbar:2006kj,Cai:2006rs,Akbar:2006er,Paranjape_2006,Nojiri:2024zdu,Nojiri:2023wzz}. In this regard, the Bekenstein-Hawking like entropy of the apparent horizon leads to the usual FLRW equations of Einstein gravity from the first law of thermodynamics of the horizon \cite{Nojiri:2024zdu,Nojiri:2023wzz}. Consequently, the natural validation of the second law of thermodynamics in Einstein gravity (cosmology) has been proposed in \cite{Odintsov:2024ipb}. Beside the Einstein gravity, the correspondence between modified theories of gravity and the horizon thermodynamics is of utmost important in order to establish a firm connection between gravity and thermodynamics. However the form of the horizon entropy corresponding to a modified theory of gravity is still questionable.

The growing interests on the interconnection between cosmology and thermodynamics leads to several proposals regarding the form of the entropy of the apparent horizon, such as, the Bekenstein-Hawking like entropy \cite{Jacobson_1995}, the Tsallis entropy \cite{Tsallis:1987eu}, the R\'{e}nyi entropy \cite{Renyi}, the Barrow entropy \cite{Barrow_2020}, the Kaniadakis entropy \cite{Kaniadakis_2005}, the Sharma-Mittal entropy \cite{Sayahian_Jahromi_2018}, the Loop Quantum gravity entropy \cite{Majhi:2017zao}, or more generally, the few parameter dependent generalized entropy \cite{Nojiri:2022aof,Nojiri_2022,Odintsov:2022qnn}. With such different forms of horizon entropies, the entropic cosmology turns out to have a rich implications on describing inflation, dark energy, reheating and bouncing as well \cite{Bousso_2002, fischler1998, Saridakis:2020zol, Sinha_2020, Adhikary:2021,Nojiri:2022aof,Nojiri_2022,Odintsov:2022qnn,Odintsov_2023,Cruz:2023xjp,Komatsu:2023wml}. On different perspective; some of our authors proposed a microscopic interpretation of such entropies in \cite{Nojiri:2023bom}, while the equivalent gravitational Lagrangian of the generalized entropy is given in \cite{Nojiri:2025fiu}. Despite these success, we still do not have a clear idea about the form of the horizon entropy corresponding to a $general~modified$ theory of gravity, or, what is the equivalent gravitational Lagrangian description for a given horizon entropy ? Actually, regarding the inter-connection between thermodynamics and cosmology, it is well known that the horizon entropy corresponding to Einstein gravity is given by the Bekenstein-Hawking like entropy. However a large class of gravity theory is governed by different modified gravity theories, such as, $f(R)$ theory, $f(Q)$ theory, Gauss-Bonnet theory etc. (with $R$ and $Q$ represent the Ricci scalar and the non-metricity scalar respectively); for which, the thermodynamic route (or particularly, the corresponding thermodynamic laws and the form of the horizon entropy) remains unknown. Therefore without having proper understanding of the aforementioned issues, the thermodynamic route of cosmology remains incomplete. Motivated by these, we try to address the following issues in the present paper:
\begin{itemize}
 \item What is the inter-connection (if any) between modified theories of gravity and the thermodynamics of the apparent horizon ?

 \item What is the equivalent gravitational Lagrangian description (if any) for a given horizon entropy ?
\end{itemize}
These in turn concrete the thermodynamic route of cosmology coming from a modified gravity theory. It turns out that the horizon entropy corresponding to a modified gravity theory (having more than one thermodynamic degrees of freedom) acquires the time derivative(s) of Bekenstein-Hawking entropy, as such derivatives act as independent thermodynamic variables coming from the first law of thermodynamics. In this regard, we may mention that recently Hollands, Wald and Zhang in \cite{Hollands:2024vbe} showed that the entropy of a dynamical black hole in general relativity also has a derivative term of the Bekenstein-Hawking entropy (with respect to the affine parameter of the null horizon generators, see also \cite{Visser:2024pwz}). By the current work, we may argue that the derivatives of Bekenstein-Hawking entropy can also appear in cosmological scenario in some modified theories of gravity particularly having more than one thermodynamic degrees of freedom.

\section{Thermodynamics of apparent horizon and the correspondence to Einstein gravity}
For a spatially flat FLRW metric given by,
\begin{eqnarray}
 ds^2 = -dt^2 + a^2(t)\left(dr^2 + r^2d\Omega^2\right) \ ,
 \label{metric}
\end{eqnarray}
(with $t$ is the cosmic time, $a(t)$ represents the scale factor of the universe and $d\Omega^2$ is the line element for a unit 2-sphere surface), the apparent horizon has the radius at,
\begin{eqnarray}
 R_\mathrm{h}(t) = \frac{1}{H(t)} \ .
 \label{app-hor}
\end{eqnarray}
Here $H(t) = \frac{\dot{a}}{a}$ is the Hubble parameter of the universe. It may be noted that, unlike to the black hole thermodynamics, the apparent horizon in cosmological context is dynamical in nature; in particular, $R_\mathrm{h}$ increases with the cosmic time provided the fluid inside the horizon satisfies the null energy condition. Here we would like to mention that in cosmological set-up, the apparent horizon depends on comoving observer (i.e. the apparent horizon changes by a change of comoving observer), which is unlike to the black hole set-up where there is one event horizon. However there is no reason to expect an exact similarity between ``black hole thermodynamics'' and ``cosmological thermodynamics''. In fact, they should not be similar, because the underlying symmetry of a black hole spacetime is very different than the cosmological spacetime. In particular, the black hole spacetime is spherically symmetric w.r.t. $one~single~point$ (which is identified as $r=0$ in the spherical coordinate system), while the cosmological spacetime (spatially flat, which we consider in the present work) is spherically symmetric w.r.t. $all~spatial~points$. Such two differently symmetric spacetimes should not be expected to have an exact similar thermodynamics.

In a cosmological set-up, the temperature of the apparent horizon is given by the surface gravity of the same and it is defined at a constant time hypersurface. Here it deserves mentioning that in a time-dependent cosmological scenario, where the presence of a timelike killing vector field is absent, a suitable notion of surface gravity is given by the Hayward-Kodama surface gravity \cite{10.1143/PTP.63.1217,Abreu_2010,Duary:2024ffj,Cai:2005ra,Akbar:2006kj,Cai:2006rs,Akbar:2006er,Paranjape_2006,Nojiri:2024zdu,Nojiri:2023wzz,Odintsov:2024ipb}. In particular, the Hayward-Kodama surface gravity in a spatially flat FLRW metric comes as: $\kappa = \frac{1}{2\sqrt{-h}}\partial_{a}\left(\sqrt{-h}h^{ab}\partial_{b}R\right)\big|_{R_\mathrm{h}}$ (with $h_{ab} = \mathrm{diag.}(-1,a^2)$ defines the induced metric along constant $\theta$ and constant $\varphi$, i.e. $h_{ab}$ is the induced metric along the normal of the apparent horizon) which is given by
\begin{eqnarray}
 \kappa = -\frac{1}{R_\mathrm{h}}\left(1 - \frac{\dot{R}_\mathrm{h}}{2}\right) \, .
 \label{surface gravity}
\end{eqnarray}
This is identified with the horizon temperature (see \cite{Nojiri:2024zdu} and the references therein), namely,
\begin{eqnarray}
 T = \frac{|\kappa|}{2\pi} = \frac{1}{2\pi R_\mathrm{h}}\left(1 - \frac{\dot{R}_\mathrm{h}}{2}\right) \, .
 \label{temperature}
\end{eqnarray}
It may be mentioned that the zeroth law of thermodynamics, in a cosmological set-up states that the temperature for all spatial points of the apparent horizon (at a constant time slice) are same, which is indeed valid in the present context.

With these ingredients in hand, we now briefly show that the Einstein gravity (cosmology) is indeed connected with the thermodynamics of the apparent horizon. In this regard, let us recall the cosmological field equations of Einstein gravity:
\begin{eqnarray}
 H^2&=&\frac{8\pi G}{3} \rho \, ,\nonumber\\
 \dot{H}&=&-4\pi G(\rho + p) \, ,
 \label{E-1}
\end{eqnarray}
where $(\rho, p)$ represent the energy density and the pressure of the fluid inside the horizon, respectively, which follows the energy conservation relation:
\begin{eqnarray}
 \dot{\rho} + 3H(\rho + p) = 0 \, .
 \label{conservation}
\end{eqnarray}
With a Bekenstein-Hawking like entropy of the horizon along with the horizon temperature ($T$) as of Eq.~(\ref{temperature}), the above FLRW Eq.~(\ref{E-1}) can be obtained from the thermodynamics of the apparent horizon provided that the first law of horizon thermodynamics is of the following form:
\begin{eqnarray}
 TdS_\mathrm{h} = -d(\rho V) + \frac{1}{2}(\rho - p)dV \, ,
 \label{E-2}
\end{eqnarray}
where $V = \frac{4\pi}{3}R_\mathrm{h}^3$ is the volume enclosed by the horizon and $S_\mathrm{h}$ is the horizon entropy which has the form: $S_\mathrm{h} \equiv S_\mathrm{BH} = \pi/(GH^2)$ for Einstein gravity (the suffix 'BH' stands for the 'Bekenstein-Hawking' entropy). Moreover the first and the second terms in the rhs of Eq.~(\ref{E-2}) represent the 'internal energy' and the 'work done' term respectively.

In establishing the connection between gravity and thermodynamics, we also need to identify the independent thermodynamic variable(s) for the gravity theory under consideration. For this purpose in the case of Einstein gravity, we take a differential of both sides of the expression $S_\mathrm{h} = \pi/(GH^2)$, leading to,
\begin{eqnarray}
 dS = -\frac{2\pi}{GH^3}dH \, .
 \label{E-3}
\end{eqnarray}
Being the only differential present in the rhs of Eq.~(\ref{E-3}) is $dH$ --- this clearly indicates that the independent thermodynamic variable, in the equivalent thermodynamic description of Einstein gravity, is given by: $\left\{H\right\}$. This depicts the reason why the horizon entropy corresponding to the Einstein gravity depends only on the Hubble parameter, as in this case, only $\left\{H\right\}$ acts as the independent thermodynamic variable and thus $S_\mathrm{h} = S_\mathrm{h}(H)$. Being a single variable function, the $S_\mathrm{h} = S_\mathrm{h}(H)$ becomes total differentiable and acts as a state function, which in turn supports the quantity $S_\mathrm{h}$ to be an $entropy$. This has important consequences in modified theories of gravity.

\section{Modified gravity with one thermodynamic d.o.f}

In this section we consider such modified theories of gravity which has different cosmological field equations than Einstein gravity, but the second FLRW equation does not contain more than single order (time) derivative of the Hubble parameter. Note that the presence up to the single derivative of the Hubble parameter in the second FLRW equation is similar to the case of Einstein gravity. As we will demonstrate that this class of modified theories is associated with one thermodynamic degree of freedom (d.o.f).

Before going to the general case of such modified theories, let us consider an explicit example. For instance, we may take the gravity theories based on non-metricity, in particular, the $F(Q)$ theory of gravity with an action like (see \cite{Heisenberg:2023lru} for a review on $F(Q)$ theory),
\begin{eqnarray}
 \mathcal{A} = \int d^4x\sqrt{-g}\left[\frac{F(Q)}{2\kappa^2} + \mathcal{L}_\mathrm{mat}\right] \, .
 \label{FQ-1}
\end{eqnarray}
Here $Q$ represents the non-metricity scalar. In a spatially flat, homogeneous and isotropic spacetime, the FLRW equation for the above action comes as,
\begin{eqnarray}
 \frac{F(Q)}{6} + 2H^2F'(Q) = \frac{8\pi G}{3}\rho \, ,
 \label{FQ-2}
\end{eqnarray}
with $F'(Q) = \frac{dF}{dQ}$ and recall that the non-metricity scalar in the spatially flat, homogeneous and isotropic spacetime is given by: $Q = -6H^2$. Therefore it may be noted from Eq.~(\ref{FQ-2}) that, similar to Einstein gravity, the first FLRW equation in $F(Q)$ theory does not contain any derivative of the Hubble parameter (or equivalently, the second FLRW equation acquires single derivative of $H$). Considering a power law form of $F(Q) = Q+\gamma Q^m$ (with $\gamma$ and $m$ are constants), the above equation takes the following form,
\begin{eqnarray}
 H^2 + \gamma(-6)^m\left(\frac{1-2m}{6}\right)H^{2m} = \frac{8\pi G}{3}\rho \, ,
 \label{FQ-3}
\end{eqnarray}
where we use $Q = -6H^2$. In the context of the apparent horizon thermodynamics, the energy density ($\rho$) times the volume enclosed by the apparent horizon acts as the total internal energy inside the horizon, in particular, $U = \rho V$ (with $V = \frac{4\pi}{3H^3}$ and $\rho$ is given by the FLRW equation). Owing to Eq.~(\ref{FQ-3}), a differential of the internal energy leads to
\begin{eqnarray}
 dU = -\frac{1}{2GH^4}\left[H^2 + \gamma(-6)^m\left(\frac{(2m-3)(2m-1)}{6}\right)H^{2m}\right]dH \, .
 \label{FQ-3a}
\end{eqnarray}
Being the only differential in the rhs of Eq.~(\ref{FQ-3a}) is $dH$, and since $dU$ acts as a thermodynamic state function, Eq.~(\ref{FQ-3a}) immediately points that the independent thermodynamic variable for $F(Q) = Q+\gamma Q^m$ cosmology is given by $\left\{H\right\}$; i.e. $F(Q) = Q+\gamma Q^m$ cosmology requires one thermodynamic d.o.f. to have a consistent thermodynamic description. In order to examine the possible link between $F(Q) = Q+\gamma Q^m$ and entropic cosmology, we define a quantity $S$ in the thermodynamic sector of the apparent horizon, such that
\begin{eqnarray}
 dS = -\frac{d(\rho V)}{T} + \frac{1}{2}(\rho - p)\frac{dV}{T} \, ,
 \label{FQ-4}
\end{eqnarray}
where $T$ represents the horizon temperature as of Eq.~(\ref{temperature}). In this regard, we should work on the following two points: (1) we first need to check whether $S$ is total differentiable. If so, then $S$ behaves as a state function and can act as the entropy of the horizon corresponding to the $F(Q) = Q+\gamma Q^m$ cosmology, otherwise, we have to find a total differentiable function to define the correct horizon entropy; (2) after identifying the horizon entropy, we then find the explicit form of the entropy that leads to the FLRW Eq.~(\ref{FQ-3}) from the first law of the horizon thermodynamics. By using Eqs.~(\ref{FQ-3}) and (\ref{FQ-4}), one gets the following expression of $dS$:
\begin{eqnarray}
 dS = -\frac{2\pi}{GH^3}\left[1 + \gamma(-6)^m\left(\frac{m-2m^2}{6}\right)H^{2m-2}\right]dH \, .
 \label{FQ-5}
\end{eqnarray}
Due to $\left\{H\right\}$ acts as the independent thermodynamic variable, Eq.~(\ref{FQ-5}) shows that the quantity $S$ depends only on $H$, i.e. $S = S(H)$ which is total differentiable (as a single variable function). Therefore the quantity $S$ behaves as a state function and can be treated as the horizon entropy for the $F(Q) = Q+\gamma Q^m$ cosmology, thus we may symbolize it as: $S \equiv S_\mathrm{h}$ (the suffix 'h' stands as the 'horizon' entropy). Here it may be mentioned that the Einstein cosmology as well as the $F(Q) = Q+\gamma Q^m$ cosmology share the same independent thermodynamic variable, namely $\left\{H\right\}$ --- this is due to the fact that the second FLRW equation in both the theories contain up to the single derivative of the Hubble parameter (the argument holds true for any general form of $F(Q)$ theory). Since $S$ itself is the horizon entropy, Eq.~(\ref{FQ-4}) is already designed in terms of the state function entropy, and thus Eq.~(\ref{FQ-4}) is considered to be the true thermodynamic law of the apparent horizon corresponding to $F(Q) = Q+\gamma Q^m$ cosmology, namely,
\begin{eqnarray}
 TdS_\mathrm{h} = -d(\rho V) + \frac{1}{2}(\rho - p)dV \, .
 \label{FQ-6}
\end{eqnarray}
 To determine the explicit form of $S_\mathrm{h}$, we use Eq.~(\ref{FQ-5}) which, by integrating once, yields
 \begin{eqnarray}
 S_\mathrm{h}= S_\mathrm{BH} + \gamma m\left(\frac{1-2m}{m-2}\right)\left(\frac{-6\pi}{G}\right)^{m-1}\frac{1}{S_\mathrm{BH}^{m-2}} \, ,
 \label{FQ-7}
\end{eqnarray}
where $S_\mathrm{BH}=\pi/(GH^2)$ is used to arrive at the above expression. Clearly the $S_\mathrm{h}$ of Eq.~(\ref{FQ-7}) reduces to the Bekenstein-Hawking entropy for $\gamma \rightarrow 0$, as expected. Eq.~(\ref{FQ-7}) gives the entropy of the apparent horizon that leads to Eq.~(\ref{FQ-3}), i.e. the cosmological field equations for $F(Q) = Q+\gamma Q^m$ gravity theory, directly from the first law of thermodynamics of the apparent horizon (\ref{FQ-6}). Therefore in the context of entropic cosmology, although the Einstein gravity and the $F(Q)$ theory share the same independent thermodynamic variable, i.e. $\left\{H\right\}$, their corresponding horizon entropy(ies) are different.

\subsection*{\underline{General Case}}\label{sec-M1}
Based on the above example, we now summarize the procedures in establishing the thermodynamic correspondence for the class of modified gravity theories where the second FLRW equation contains up to the first order time derivative of the Hubble parameter.
\begin{itemize}
 \item For such modified gravity theories, the FLRW equations take the following form:
\begin{eqnarray}
 F(H)&=&\frac{8\pi G}{3}\rho \, ,\nonumber\\
 \left(\frac{\partial F}{\partial H^2}\right)\dot{H}&=&-4\pi G(\rho + p) \, ,
 \label{Gen1-a}
\end{eqnarray}
where $(\rho, p)$ satisfy the conservation relation (\ref{conservation}). Moreover $F(H)$ is any analytic function of the Hubble parameter, depending on the gravity theory under consideration; for instance, the functional form of $F(H)$ for $F(Q) = Q + \gamma Q^{m}$ theory is shown in Eq.~(\ref{FQ-3}).

\item The first FLRW equation suggests that $\rho$ depends only on $H$, and thus a differential of the internal energy ($U=\rho V$: a state function in apparent horizon thermodynamics) comes as follows,
\begin{eqnarray}
 dU = \left(\rho \frac{\partial V}{\partial H} + V \frac{\partial \rho}{\partial H}\right)dH \, ,
 \label{ind-var}
\end{eqnarray}
which clearly states that the independent thermodynamic variable, in the thermodynamic description of the modified theories governed by Eq.~(\ref{Gen1-a}), is given by $\left\{H\right\}$. In order to examine the possible connection between such modified gravity and the horizon thermodynamics, we define a quantity $S$ in the thermodynamic sector by the way:
\begin{eqnarray}
 dS = -\frac{d(\rho V)}{T} + \frac{1}{2}(\rho - p)\frac{dV}{T} \, ,
 \label{Gen1-b}
\end{eqnarray}
with $T$ being the temperature of the horizon. At this stage, it is important to check whether $S$ behaves as a state function and can stand as the entropy of the horizon. If so, we then can express the thermodynamic law of the apparent horizon in terms of the correct horizon entropy.

\item For this purpose, one may use Eqs.~(\ref{Gen1-a}) and (\ref{Gen1-b}) to get,
\begin{eqnarray}
 dS = -\frac{\pi}{GH^4}\left(\frac{\partial F}{\partial H}\right)dH \, ,
 \label{Gen1-c}
\end{eqnarray}
depicting that $S$ depends on the independent variable $H$, i.e. $S= S(H)$. Therefore, being a single variable function, $S$ becomes total differentiable (or a state function) and thus can act as the horizon entropy for the present case.

\item Once $S$ is identified to be the horizon entropy (i.e. $S \equiv S_\mathrm{h}$), the thermodynamic law of the apparent horizon, corresponding to the present class of modified theories of gravity, can be immediately written from Eq.~(\ref{Gen1-b}), and is given by,
\begin{eqnarray}
 TdS_\mathrm{h} = -d(\rho V) + \frac{1}{2}(\rho - p)dV \, .
 \label{Gen1-d}
\end{eqnarray}

\item Finally the explicit form of $S_\mathrm{h}$ can be obtained by integrating Eq.~(\ref{Gen1-c}), which yields
\begin{eqnarray}
 S_\mathrm{h} = -\frac{\pi}{G}\int \frac{1}{H^4}\left(\frac{\partial F}{\partial H}\right)dH\bigg|_{S_\mathrm{BH} = \pi/(GH^2)} \, ,
 \label{Gen1-e}
\end{eqnarray}
which can be expressed in terms of the $S_\mathrm{BH} = \pi/(GH^2)$. For Einstein gravity, $F(H) = H^2$, and thus the $S_\mathrm{h}$ of Eq.~(\ref{Gen1-e}) reduces to the Bekenstein-Hawking form; while for other gravity theories where $F(H) \neq H^2$, the horizon entropy acquires a different form than the Bekenstein-Hawking one. Therefore in the equivalent thermodynamic description of the modified gravity theories governed by Eq.~(\ref{Gen1-a}) --- the independent thermodynamic variable is given by $\left\{H\right\}$ (evident from Eq.~(\ref{ind-var})), while the thermodynamic law and the horizon entropy are shown in Eqs.~(\ref{Gen1-d}) and (\ref{Gen1-e}) respectively.
\end{itemize}

\section{Modified gravity with more than one thermodynamic d.o.f}
We now focus to the modified gravity theories where the second FLRW equation contains higher time derivatives of the Hubble parameter (see \cite{Nojiri:2010wj,Capozziello:2011et} for an extended review on modified gravity). It will turn out that this class of modified theory requires more than one thermodynamic degrees of freedom (d.o.f) in order to have their consistent thermodynamic description. However before moving to a general form of such modified gravity, let us work out with two definite examples, for instance $F(R) = R+ \alpha R^2$ and $F(R,\mathcal{G}) = R+\beta \mathcal{G}^2$ respectively (where $R$ is the Ricci scalar and $\mathcal{G}$ is the Gauss-Bonnet curvature term).

\subsection*{\underline{An example: $F(R) = R+\alpha R^2$ and entropic cosmology}}\label{sec-FR}
With an action for $F(R)$ gravity like \cite{Nojiri:2010wj,Capozziello:2011et},
\begin{eqnarray}
 \mathcal{A} = \int d^4x\sqrt{-g} \left[\frac{F(R)}{16\pi G} + \mathcal{L}_\mathrm{mat}\right] \, ,
 \label{F-1}
\end{eqnarray}
along with a spatially flat, homogeneous and isotropic metric (\ref{metric}), the FLRW equations for the $F(R)$ theory are given by,
\begin{eqnarray}
 \frac{F(R)}{2}-3\left(H^2 + \dot{H}\right)F'(R) + 18\left(4H^2\dot{H} + H\ddot{H}\right)F''(R) = (8\pi G)\rho \, ,
 \label{F-1a}
 \end{eqnarray}
 {\small{\begin{eqnarray}
 \frac{F(R)}{2} - \left(\dot{H} + 3H^2\right)F'(R) + 6\left(8H^2\dot{H} + 4\dot{H}^2 + 6H\ddot{H} + \dddot{H}\right)F''(R)&+&36\left(4H\dot{H} + \ddot{H}\right)^2F'''(R)\nonumber\\
 &=&-(8\pi G)p\, ,
 \label{F-2}
\end{eqnarray}}}
where $G$ is the Newton's gravitational constant, $F'(R) = \frac{dF}{dR}$, and $(\rho,p)$ denote the energy density and the pressure of the fluid inside the horizon, respectively. The above two equations immediately lead to the local conservation relation of the fluid, namely Eq.~(\ref{conservation}). Therefore the two FLRW equations and the conservation equation are not independent to each other, actually one of these can be determined from the other two. We consider the first FLRW equation and the conservation relation to be the independent ones. For $F(R) = R+\alpha R^2$, the first FLRW Eq.~(\ref{F-1a}) takes the following form,
\begin{eqnarray}
 H^2 + 6\alpha\left(6H^2\dot{H} + 2H\ddot{H} - \dot{H}^2\right) = \frac{8\pi G}{3}\rho\, ,
 \label{F-4}
\end{eqnarray}
where an overdot $\equiv \frac{d}{dt}$), and consequently, the second FLRW equation can also be obtained from Eq.~(\ref{F-2}) which contains up to third derivative of the Hubble parameter. The FLRW Eq.~(\ref{F-4}) shows that $\rho = \rho(H,\dot{H},\ddot{H})$, and thus the differential of the internal energy $U = \rho V$ (which is a state function in the context of horizon thermodynamics) turns out to be,
\begin{eqnarray}
 dU = \left(\rho \frac{\partial V}{\partial H} + V \frac{\partial \rho}{\partial H}\right)dH + V \frac{\partial \rho}{\partial \dot{H}}d\dot{H} + V \frac{\partial \rho}{\partial \ddot{H}}d\ddot{H} \, .
 \label{F-4a}
\end{eqnarray}
This confirms that the thermodynamic scenario, corresponding to $F(R) = R+\alpha R^2$ cosmology, treats $\left\{H, \dot{H}, \ddot{H}\right\}$ to be the independent thermodynamic variables; and therefore the other thermodynamic quantities are determined in terms of these three variables. It may be noted from Eq.~(\ref{F-4a}) that for $\alpha = 0$, all the coefficients of $d\dot{H}$ and $d\ddot{H}$ vanish, and thus only $\left\{ H \right\}$ acts as the independent thermodynamic variable in the case of Einstein gravity --- this is consistent with the fact that the correct horizon entropy for Einstein gravity depends only on $H$. In order to examine the possible link between $F(R) = R+\alpha R^2$ and entropic cosmology, we first need to identify the entropy of the apparent horizon, which should be a state function. For this purpose, let us define a quantity $S$ in the thermodynamic sector of the apparent horizon by the way:
\begin{eqnarray}
 dS = -\frac{d(\rho V)}{T} + \frac{1}{2}(\rho - p)\frac{dV}{T} \, ,
 \label{F-5}
\end{eqnarray}
with $T$ is the horizon temperature as of Eq.~(\ref{temperature}). Similar to the previous cases, we now need to check whether $S$ is a state function and can be the correct entropy of the horizon (in this regard, we will get considerable differences compared to the earlier cases). By using Eq.~(\ref{F-4}), one easily gets the differential of $S$ from Eq.~(\ref{F-5}) as follows:
\begin{eqnarray}
 dS = -\left(\frac{\pi}{G}\right)\left(\frac{2}{H^3} + \frac{72\alpha\dot{H}}{H^3} + \frac{12\alpha\ddot{H}}{H^4}\right)dH + \left(\frac{36\alpha}{H^2} - \frac{12\alpha\dot{H}}{H^4}\right)d\dot{H} + \left(\frac{12\alpha}{H^3}\right)d\ddot{H}\, .
 \label{F-7}
\end{eqnarray}
The following important points deserves mentioning at this stage:
\begin{enumerate}
 \item Being $\left\{H, \dot{H}, \ddot{H}\right\}$ are the independent variables, the entity $S$ is not total differentiable. Therefore $S$ is not a state function and can not be the true thermodynamic entropy for $F(R) = R+\alpha R^2$ cosmology. This is a generic problem for the modified gravity theories, particularly where the (second) FLRW equation contains higher derivatives of the Hubble parameter and have more than one thermodynamic d.o.f., as we will show at the end of this section. Here we would like to mention that such an issue does not arise in the modified theories discussed in Sec.~[\ref{sec-M1}] because, in this case, the independent variable is given by only $\left\{H\right\}$ and thus the $S=S(H)$ becomes total differentiable (as a single variabe function) which safely acts as the corresponding thermodynamic entropy.

\item We may note from Eq.~(\ref{F-7}) that, rather than $S$, the quantity which is total differentiable is given by:
\begin{eqnarray}
 -\left(\frac{\pi}{G}\right)\frac{dS}{S_\mathrm{BH}^2} = \left(2H + 72\alpha H\dot{H} + 12\alpha \ddot{H}\right)dH + \left(36\alpha H^2 - 12\alpha\dot{H}\right)d\dot{H} + 12\alpha H d\ddot{H} \, ,
 \label{F-7a}
\end{eqnarray}
i.e. $\frac{1}{S_\mathrm{BH}^2}$ plays the role of the integrating factor. Consequently we may express,
\begin{eqnarray}
 \frac{dS}{S_\mathrm{BH}^2} = -d\left(\frac{1}{S_\mathrm{h}}\right) \, ,
 \label{F-7b}
\end{eqnarray}
where $S_\mathrm{h}$ is total differentiable, and as a result, Eq.~(\ref{F-7a}) can be written as,
\begin{eqnarray}
 -\left(\frac{\pi}{G}\right)\frac{dS_\mathrm{h}}{S_\mathrm{h}^2} = \left(2H + 72\alpha H\dot{H} + 12\alpha \ddot{H}\right)dH + \left(36\alpha H^2 - 12\alpha\dot{H}\right)d\dot{H} + 12\alpha H d\ddot{H} \, .
 \label{F-7b1}
\end{eqnarray}
In terms of the state function $S_\mathrm{h}$, Eq.~(\ref{F-5}) takes the following form:
\begin{eqnarray}
 TdS_\mathrm{h} = -d\left(\rho V\right) + \left\{\frac{1}{2}(\rho - p) + \frac{1}{2}\left[1 - \left(\frac{3H^2}{8\pi G\rho}\right)^2\right](\rho + p)\left(1 + \frac{2H^2}{\dot{H}}\right)\right\}dV \, ,
 \label{F-7c}
\end{eqnarray}
with $\rho = \rho(H,\dot{H},\ddot{H})$ is shown in Eq.~(\ref{F-4}). In the limit of Einstein gravity --- $S_\mathrm{h}$ takes the Bekenstein-Hawking form (see Eq.~(\ref{F-7b})) and $H^2 = \frac{8\pi G\rho}{3}$, then Eq.~(\ref{F-7c}) becomes $TdS_\mathrm{BH} = -d(\rho V) + \frac{1}{2}(\rho - p)dV$ which is the thermodynamic law of the apparent horizon for Einstein gravity. Therefore we may argue that Eq.~(\ref{F-7c}) represents the thermodynamic law of the apparent horizon corresponding to $F(R) = R+\alpha R^2$ cosmology with the identifications like: $T$ is the temperature of the horizon (coming from the surface gravity of the horizon); $S_\mathrm{h}$ is the horizon entropy (a state function); the first term in the rhs is the change of internal energy; while the second term in the rhs, which is proportional to $dV$, shows as work done. Moreover, in support for the quantity $S_\mathrm{h}$ to be an $entropy$, let us use Eqs.~(\ref{F-5}) and (\ref{F-7b}) to get
\begin{eqnarray}
 \frac{dS_\mathrm{h}}{S_\mathrm{h}^2} = -\frac{8G^2}{3}d\rho \, ,
 \label{F-7d}
\end{eqnarray}
where the conservation relation of the fluid has been used. The above expression clearly shows the monotonic increasing behaviour of $S_\mathrm{h}$ with the cosmic time (as long as the fluid satisfies the null energy condition) --- this actually supports the increasing behaviour of an entropy function. Later we will show that, beside the $F(R) = R+\alpha R^2$ cosmology, the thermodynamic law for $general$ modified gravity theories (cosmology) is given by the same form as of Eq.~(\ref{F-7c}) which also has a limit to the Einstein gravity.
\end{enumerate}

Having obtained the thermodynamic law in Eq.~(\ref{F-7c}), we now intend to find the form of horizon entropy ($S_\mathrm{h}$) corresponding to the $F(R)=R+\alpha R^2$ cosmology. For this purpose, we now integrate Eq.~(\ref{F-7b1}) to obtain the following expression of $S_\mathrm{h}$ as,
\begin{eqnarray}
 S_\mathrm{h} = \frac{\pi}{GH^2}\left[1 + 6\alpha\left(6\dot{H} + 2\frac{\ddot{H}}{H} - \frac{\dot{H}^2}{H^2}\right)\right]^{-1} \, .
 \label{F-7f}
\end{eqnarray}
With the help of $S_\mathrm{BH} = \pi/(GH^2)$, the rhs of the above equation can be expressed in terms of $\left\{S_\mathrm{BH}, \dot{S}_\mathrm{BH}, \ddot{S}_\mathrm{BH}\right\}$, and by doing so, we get
\begin{eqnarray}
 S_\mathrm{h} = S_\mathrm{BH}\left[1 - 18\alpha\left(\frac{\pi}{G}\right)^{1/2}\frac{\dot{S}_\mathrm{BH}}{S_\mathrm{BH}^{3/2}} + \frac{15\alpha}{2}\frac{\dot{S}_\mathrm{BH}^2}{S_\mathrm{BH}^2} - 6\alpha\frac{\ddot{S}_\mathrm{BH}}{S_\mathrm{BH}}\right]^{-1} \, .
 \label{F-7g}
\end{eqnarray}
Eq.~(\ref{F-7g}) gives the entropy of the apparent horizon that leads to Eq.~(\ref{F-4}), i.e. the cosmological field equations for $F(R)=R+\alpha R^2$ gravity theory, directly from the first law of thermodynamics (\ref{F-7c}) of the apparent horizon. In the correspondence between $F(R)=R+\alpha R^2$ and entropic cosmology, it turns out that $\left\{S_\mathrm{BH}, \dot{S}_\mathrm{BH}, \ddot{S}_\mathrm{BH}\right\}$ act as the independent thermodynamic variables (as evident from Eq.~(\ref{F-4a})), and thus the horizon entropy depends on these three variables (recall that the independent variables may be given by $\left\{H, \dot{H}, \ddot{H}\right\}$ or $\left\{S_\mathrm{BH}, \dot{S}_\mathrm{BH}, \ddot{S}_\mathrm{BH}\right\}$ due to $S_\mathrm{BH} = \frac{\pi}{GH^2}$). The other thermodynamic state quantities, for instance the internal energy, can also be identified in terms of these independent thermodynamic variables (see Eq.~(\ref{M-8}) where we show such identifications in general modified gravity). Moreover it may be noted from Eq.~(\ref{F-7g}) that the coefficients of the time derivatives of $S_\mathrm{BH}$ in the expression of $S_\mathrm{h}$ are proportional to the higher curvature parameter $\alpha$, which identically vanish for Einstein gravity.

Since the thermodynamic system (corresponding to $F(R)=R+\alpha R^2$ cosmology) has three d.o.f., the thermodynamic state of the system may be represented by a 3-dimensional configuration space where the coordinates of a point is given by $\equiv (H,\dot{H},\ddot{H})$. Any two arbitrary points in the configuration space are connected by many possible paths, including the classical one governed by the FLRW Eq.~(\ref{F-4}) where the boundary conditions are given by the coordinates of the considered points. In this regard, we would like to mention that Eq.~(\ref{F-7b1}) (or equivalently, Eq.~(\ref{F-7f})) is valid for all the possible configurations, in particular, Eq.~(\ref{F-7b1}) can be used to calculate the change of entropy between two points in the configuration space via any possible paths (connecting the two points). On other hand, the thermodynamic law (\ref{F-7c}) (in terms of $dU$ and $dV$) is obtained by using the FLRW Eq.~(\ref{F-4}). This indicates that, unlike to Eq.~(\ref{F-7b1}), the thermodynamic law (\ref{F-7c}) is defined over only the classical path between two points in the configuration space.

The thermodynamic correspondence of $F(R,\mathcal{G}) = \frac{R}{16\pi G} + \beta \mathcal{G}^2$ theory and entropic cosmology is discussed in the Appendix- Sec.~[\ref{sec-appendix}].

\subsection*{\underline{General Case}}\label{sec-M2}
Based on these examples, we now summarize the thermodynamic equivalence of the class of modified gravity theories where the second FLRW equation contains higher derivative(s) of the Hubble parameter, and the demonstration goes as follows:
\begin{itemize}
 \item For this class of modified gravity, the FLRW equation for spatially flat, homogeneous and isotropic spacetime can be written as \cite{Nojiri:2010wj,Capozziello:2011et},
 \begin{eqnarray}
  F\left(H, \dot{H}, \ddot{H},....., H^{(n-1)}\right) = \frac{8\pi G}{3}\rho \, ,
  \label{M-1}
 \end{eqnarray}
where the lhs of Eq.~(\ref{M-1}) represents a function of the Hubble parameter and its time derivatives (with $H^{(n-1)} = \frac{d^{n-1}H}{dt^{n-1}}$) and the explicit functional form depends on the gravity theory under consideration. For instance, in $R+\alpha R^2$ theory, $n=3$ and the functional form is shown in the lhs of Eq.~(\ref{F-4}). Here it may be mentioned that Eq.~(\ref{M-1}) along with the conservation of the fluid leads to the second FLRW equation, and since the first FLRW Eq.~(\ref{M-1}) contains up to $(n-1)$-th order time derivative of the Hubble parameter, the second FLRW equation acquires up to $n$-th order derivative of $H$.

\item Owing to the FLRW Eq.~(\ref{M-1}), the differential of internal energy ($U = \rho V$) comes as,
\begin{eqnarray}
 dU = \frac{3}{8\pi G}\left[\left(V\frac{\partial F}{\partial H} + F\frac{\partial V}{\partial H}\right)dH + V\sum_{i=1}^{n-1}\frac{\partial F}{\partial H^{(i)}}dH^{(i)}\right] \, .
 \label{M-1a}
\end{eqnarray}
Therefore in the context of apparent horizon thermodynamics where $dU$ is a state function, Eq.~(\ref{M-1a}) provides the independent thermodynamic variable corresponding to the modified theories (governed by Eq.~(\ref{M-1})) as: $\left\{H, \dot{H}, \ddot{H},...., H^{(n-1)}\right\}$. In order to examine the possible link between the cosmology of such modified theories and the entropic cosmology, we define a quantity $S$ in the thermodynamic sector, via,
\begin{eqnarray}
 dS = -\frac{d(\rho V)}{T} + \frac{1}{2}(\rho - p)\frac{dV}{T} \, ,
 \label{M-2}
\end{eqnarray}
with $T$ is given in Eq.~(\ref{temperature}). Such a quantity, namely $S$, turns out to be important to define the correct entropy of the apparent horizon corresponding to the present class of modified theories. By using the FLRW Eq.~(\ref{M-1}), one gets
\begin{eqnarray}
 dS = -\left(\frac{8\pi^2}{3H^4}\right)\sum_{i=0}^{n-1}\frac{\partial F}{\partial H^{(i)}}dH^{(i)} \, .
 \label{M-3}
\end{eqnarray}
Here the function $F(H, \dot{H}, \ddot{H},....., H^{(n-1)})$ is a total differential as it is connected with the energy density via the FLRW equation; however due to the presence of the factor $\sim \frac{1}{H^4}$ in front of the summation, the quantity $S$ fails to be a total differentiable function. As a result, $S$ can not directly act as the entropy of the apparent horizon. However Eq.~(\ref{M-3}) also suggests that a factor like $\sim H^4$ can be taken as an integrating factor, in particular,
\begin{eqnarray}
 \frac{dS}{S_\mathrm{BH}^2} \sim \sum_{i=0}^{n-1}\frac{\partial F}{\partial H^{(i)}}dH^{(i)} \, ,
 \label{New-1}
\end{eqnarray}
is total differentiable. Therefore we may write,
\begin{eqnarray}
 \frac{dS}{S_\mathrm{BH}^2} = -d\left(\frac{1}{S_\mathrm{h}}\right) \, .
 \label{M-4}
\end{eqnarray}
Here $S_\mathrm{h}$ is total differentiable, which, due to Eq.~(\ref{New-1}), comes as,
\begin{eqnarray}
 \frac{dS_\mathrm{h}}{S_\mathrm{h}^2} = -\left(\frac{\pi}{G}\right) \sum_{i=0}^{n-1}\frac{\partial F}{\partial H^{(i)}}dH^{(i)} \, .
 \label{New-2}
\end{eqnarray}
Moreover, Eqs.~(\ref{M-3}) and (\ref{M-4}), immediately yield to,
\begin{eqnarray}
 \frac{dS_\mathrm{h}}{S_\mathrm{h}^2} = -\frac{8G^2}{3}d\rho \, ,
 \label{M-5}
\end{eqnarray}
which depicts an important fact that $S_\mathrm{h}$ is a monotonic increasing function of time provided the fluid obeys the null energy condition. In the present work we indeed consider the validity of the null energy condition (i.e. $\rho + p > 0$), and thus $S_\mathrm{h}$ increases with the cosmic time.

\item In terms of the state function $S_\mathrm{h}$, Eq.~(\ref{M-2}) can be expressed as,
\begin{eqnarray}
 TdS_\mathrm{h} = -d\left(\rho V\right) + \left\{\frac{1}{2}(\rho - p) + \frac{1}{2}\left[1 - \left(\frac{3H^2}{8\pi G\rho}\right)^2\right](\rho + p)\left(1 + \frac{2H^2}{\dot{H}}\right)\right\}dV \, ,
 \label{M-6}
\end{eqnarray}
where we use Eq.~(\ref{M-5}). In the limit of Einstein gravity, the above expression reduces to the correct form of thermodynamic law as of Eq.~(\ref{E-2}). Therefore we may argue that the expression of Eq.~(\ref{M-6}) represents the thermodynamic law for the present class of modified gravity theories (governed by Eq.~(\ref{M-1})) with the following identifications: $T$ is the horizon temperature, $S_\mathrm{h}$ is the horizon entropy (being a state function and monotonically increasing with time), and the terms in the rhs represent the internal energy and the work done (proportional to $dV$) term respectively.

\item Regarding the explicit form of the horizon entropy, we use Eq.~(\ref{New-2}) which can be integrated once to arrive at the following expression of $S_\mathrm{h}$:
\begin{eqnarray}
 S_\mathrm{h} = \left(\frac{\pi}{G}\right)F^{-1}(H, \dot{H}, \ddot{H},....., H^{(n-1)}) \, ,
 \label{M-7}
\end{eqnarray}
or equivalently, this can be expressed in terms of $S_\mathrm{BH} = \pi/(GH^2)$ and its derivatives. Hence the horizon entropy corresponding to the present class of modified gravity theories depends on $S_\mathrm{BH}$ and its derivatives up to $(n-1)$-th order. This is however expected from the fact that $\left\{H, \dot{H}, \ddot{H},...., H^{(n-1)}\right\}$ act as the independent thermodynamic variables for such modified theories, and consequently, the other thermodynamic quantities (including the horizon entropy) are supposed to depend on these independent variables. Similarly the internal energy, in terms of these independent variables, can be expressed as,
\begin{eqnarray}
 dU=d(\rho V) = \frac{1}{2GH^3}\left[\left(\frac{\partial F}{\partial H} - \frac{3F}{H}\right)dH + \sum_{i=1}^{n-1} \frac{\partial F}{\partial H^{(i)}}dH^{(i)}\right] \, ,
 \label{M-8}
\end{eqnarray}
where the function $F(H, \dot{H}, \ddot{H},....., H^{(n-1)})$ represents the FLRW Eq.~(\ref{M-1}). Thus as a whole, the thermodynamic law and the horizon entropy corresponding to the modified theories (with the FLRW Eq.(\ref{M-1})) are shown in Eqs.~(\ref{M-6}) and (\ref{M-7}) respectively.
\end{itemize}

\section{Second law of thermodynamics in modified theories of gravity}
 Having obtained the first law of thermodynamics (and the corresponding form of horizon entropy), we now focus to the second law of thermodynamics for ``general'' modified theories of gravity, which is important from its own right. In this regard, we need to evaluate the change of total entropy, i.e. the horizon entropy ($S_\mathrm{h}$) and the entropy of the matter fields inside the horizon ($S_\mathrm{m}$), w.r.t. the cosmic expansion of the universe. Here we consider the class of modified gravity theories having one thermodynamic d.o.f, which are governed by Eq.~(\ref{Gen1-a}); however the generalization for more than one thermodynamic d.o.f. can be similarly formulated.

 Owing to Eq.~(\ref{Gen1-a}) and Eq.~(\ref{Gen1-c}), we have
 \begin{eqnarray}
  dS_\mathrm{h} \sim -\left(\frac{1}{H^4}\right)d\rho \, ,
  \label{s-1}
 \end{eqnarray}
which shows that the horizon entropy monotonically increases with cosmic time ($t$), provided the matter fields obey the null energy condition, i.e. $\omega > -1$. In the present work, we will not consider any phantom fields, in particular, the EoS of the matter fields satisfies $\omega > -1$.

\begin{figure}[!h]
\begin{center}
\centering
\includegraphics[width=2.0in,height=2.0in]{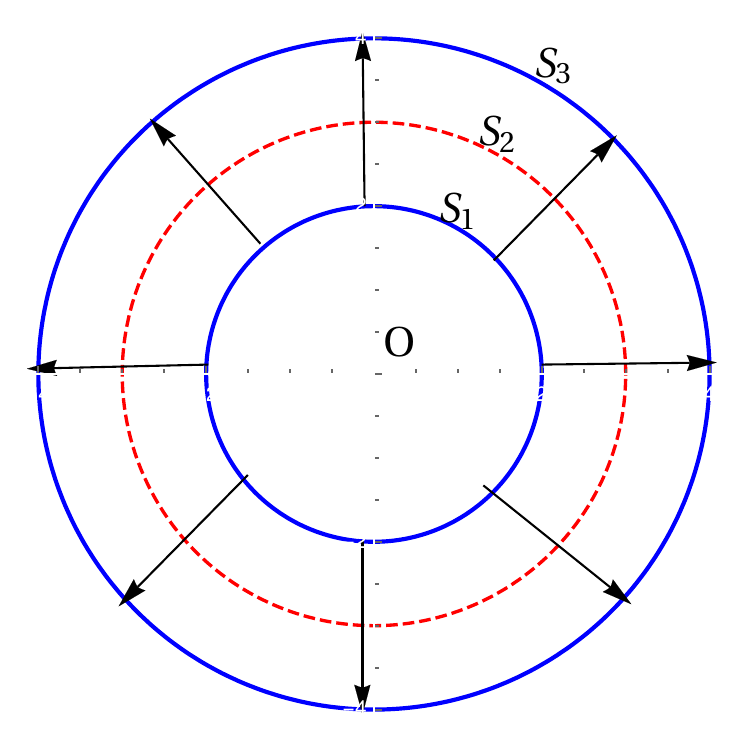}
\caption{Comparison between the formation of apparent horizon and the comoving expansion of the universe, in order to calculate the matter flux through the horizon.}
 \label{plot-2}
\end{center}
\end{figure}

Regarding the matter fields inside the horizon, we consider the matter fields to be a perfect fluid with a $constant$ equation of state (EoS) parameter $\omega$ given by:
\begin{eqnarray}
 p = \omega \rho \, ,
 \label{new-3}
\end{eqnarray}
where $p$ and $\rho$ represent the pressure and the energy density of the matter field, respectively. Depending on the values of $\omega$, the universe undergoes through different cosmic stages. In the present work, we will use a general $\omega$ without putting any constraint on it. Our motive is to investigate which of the cosmic era (naturally) allows the irreversibility or the reversibility of second law of thermodynamics in the present context.\\
The matter fields behaves like an open system as it exhibits a flux through the apparent horizon. Such kind of matter flux exists due to the the difference between the comoving expansion speed of the universe ($v_\mathrm{c}$) and the speed of the formation of the apparent horizon ($v_\mathrm{h}$). In particular, $v_\mathrm{c} = HD$ (at a physical distance $D$ from a comoving observer) and $v_\mathrm{h} = -\dot{H}/H^2$. Therefore the thermodynamic law of the matter fields inside the horizon can be expressed by,
\begin{eqnarray}
 T_\mathrm{m}dS_\mathrm{m}&=&(\mathrm{increase~of~internal~energy}) + (\mathrm{work~done}) + (\mathrm{energy~flux~through~horizon}) \, ,
 \label{thermo-m-1}
\end{eqnarray}
where $T_\mathrm{m}$ and $S_\mathrm{m}$ represent the temperature and entropy of the matter fields respectively. In general, the matter fields' temperature is considered to be different than the horizon temperature --- we will explicitly examine this issue at some stage. The internal energy of the matter fields (at instant $t$) is given by $E(t) = \rho(t)V(t)$ and thus the increase of internal energy during the cosmic interval $dt$ becomes (by using the conservation law of matter field),
\begin{eqnarray}
 dE = -3H\left(\rho + p\right)Vdt + \rho dV \, .
 \label{thermo-m-2}
\end{eqnarray}
Regarding the second term in Eq.~(\ref{thermo-m-1}), the work done by the matter fields is expressed as,
\begin{eqnarray}
 dW = \frac{1}{2}T_\mathrm{ab}h^{ab}dV = \frac{1}{2}\left(p - \rho\right)dV \, ,
 \label{thermo-m-3}
\end{eqnarray}
here the work density is defined by the projection of the energy-momentum tensor of the matter fields along the normal of the apparent horizon \cite{Cai:2005ra,Cai:2006rs}, where $h^{ab}$ is the induced metric along the normal of the apparent horizon. For the third term in Eq.~(\ref{thermo-m-1}), we need to realize that the matter fields exhibit a flux through the horizon due to $v_\mathrm{c} \neq v_\mathrm{h}$, and the demonstration is shown in Fig.~[\ref{plot-2}] where the concentric spheres (with respect to the comoving observer labeled by 'O') represent as follows --- (a) $S_\mathrm{1}$: the visible universe bounded by the apparent horizon at time $t$, having radius $OS_\mathrm{1} = 1/H(t)$; (b) $S_\mathrm{2}$: the visible universe bounded by the apparent horizon at time $t+dt$, having radius $OS_\mathrm{2} = 1/H(t+dt) = \frac{1}{H}-\frac{\dot{H}}{H^2}dt$ (at the leading order in $dt$); (c) $S_\mathrm{3}$: due to the difference between $v_\mathrm{c}$ and $v_\mathrm{h}$, the sphere $S_\mathrm{1}$ moves from $S_\mathrm{1} \rightarrow S_\mathrm{3}$ due to the comoving expansion of the universe and thus the radius of $S_\mathrm{3}$ turns out to be $OS_\mathrm{3} = \frac{1}{H} + dt$ (as $v_\mathrm{c}(t) = 1$ on the apparent horizon).
Thereby we calculate, the outward flux of the matter fields' energy through the horizon as,
\begin{eqnarray}
 \mathrm{Flux}=\rho\times\left[V_\mathrm{c}(t+dt) - V(t+dt)\right] = \frac{4\pi\rho}{H^2}\left(1 + \frac{\dot{H}}{H^2}\right)dt
 = -\frac{2\pi\rho}{H^2}\left(1 + 3\omega_\mathrm{eff}\right)dt \, ,
 \label{thermo-m-5}
\end{eqnarray}
in the leading order of $dt$ and $\omega_\mathrm{eff} = -1-2\dot{H}/(3H^2)$. Recall that in modified theories of gravity (other than the Einstein gravity), the $\omega_\mathrm{eff}$ and the matter fields' EoS ($\omega$, see Eq.~(\ref{new-3})) are different. Owing to Eqs.~(\ref{thermo-m-2}), (\ref{thermo-m-3}) and (\ref{thermo-m-5}), the thermodynamic law for the matter fields inside the horizon from Eq.~(\ref{thermo-m-1}) turns out to be,
\begin{eqnarray}
 T_\mathrm{m}\frac{dS_\mathrm{m}}{dt} = -\frac{\pi\rho}{H^2}(1-3\omega_\mathrm{eff})\left((1+\omega) + \frac{2(1+3\omega_\mathrm{eff})}{(1-3\omega_\mathrm{eff})}\right) \, .
 \label{thermo-m-6}
\end{eqnarray}
This clearly shows that $T_\mathrm{m}\dot{S}_\mathrm{m} < 0$ in the range $\omega > -1$ of our interest, i.e. the entropy of the matter fields proves to monotonically decrease with the cosmic time.

From the above discussions, particularly from Eq.~(\ref{s-1}) and Eq.~(\ref{thermo-m-6}), it becomes clear that the entropy of the horizon increases while the matter fields' entropy decreases with the time. Therefore we have,
\begin{eqnarray}
 T_\mathrm{h}\frac{dS_\mathrm{h}}{dt} > 0~~~~~\mathrm{and}~~~~~T_\mathrm{m}\frac{dS_\mathrm{m}}{dt} < 0 \, ,
 \label{exc-1}
\end{eqnarray}
which reveal that the heat energy is released by the matter fields and is absorbed by the apparent horizon. This immediately indicates that the flow of heat energy is directed from the matter fields to the apparent horizon during the cosmic expansion of the universe. Such spontaneous direction of heat flow in turn points the following inequality:
\begin{eqnarray}
 T_\mathrm{m} \geq T_\mathrm{h} \, .
 \label{exc-2}
\end{eqnarray}
This opens two different possibilities: $T_\mathrm{m} = T_\mathrm{h}$ or $T_\mathrm{m} > T_\mathrm{h}$. In the following, we examine which of these two possibilities are allowed by different cosmic era of the universe.

For the case: $T_\mathrm{m} = T_\mathrm{h}$, the matter fields should be in thermal equilibrium with the horizon and the heat flow from the matter fields $\rightarrow$ the horizon is reversible in nature, i.e.,
\begin{eqnarray}
 \Delta S_\mathrm{h} + \Delta S_\mathrm{m} = 0 \, .
 \label{sub-1-1}
\end{eqnarray}
For the other case: $T_\mathrm{m} > T_\mathrm{h}$, the matter fields are not in thermal equilibrium with the horizon and the heat exchange from the matter fields to the horizon should be irreversible in nature where
\begin{eqnarray}
 \Delta S_\mathrm{h} + \Delta S_\mathrm{m} > 0 \, .
 \label{sub-2-1}
\end{eqnarray}
If $\left|\Delta Q_\mathrm{m}\right|$ is the amount of heat released by the matter fields (within the interval when the horizon entropy increases by $\Delta S_\mathrm{h}$), then
\begin{eqnarray}
 \left|\Delta Q_\mathrm{m}\right| = T_\mathrm{m}\left|\Delta S_\mathrm{m}\right| \, ,
 \label{sub-1-2}
\end{eqnarray}
which, due to $T_\mathrm{m} \geq T_\mathrm{h}$, takes the following form:
\begin{eqnarray}
 \left|\Delta Q_\mathrm{m}\right| \geq T_\mathrm{h}\Delta S_\mathrm{h}\left(\frac{\left|\Delta S_\mathrm{m}\right|}{\Delta S_\mathrm{h}}\right) \, ,
 \label{sub-1-3}
\end{eqnarray}
where the equality (or inequality) sign designate the reversible (or irreversible) flow of heat energy from the matter fields to the horizon (recall, $\Delta Q_\mathrm{m} < 0$). Due to $\Delta S_\mathrm{h} + \Delta S_\mathrm{m} \geq 0$, the above expression gets automatically satisfied if,
\begin{eqnarray}
 \left|\Delta Q_\mathrm{m}\right| \geq T_\mathrm{h}\Delta S_\mathrm{h} \, .
 \label{sub-2-4}
\end{eqnarray}
Owing to Eq.~(\ref{s-1}) and Eq.~(\ref{thermo-m-6}), the above expression leads to,
\begin{eqnarray}
 1+3\omega_\mathrm{eff} \geq 0 \, .
 \label{s-3}
\end{eqnarray}
Therefore the reversible or irreversible case(s) of second law of thermodynamics, in modified gravity (cosmology), demands the above condition (on $\omega_\mathrm{eff}$) to be hold. This interestingly demonstrates the following points: (1) the irreversible case of second law of thermodynamics is allowed for $\omega_\mathrm{eff} > -\frac{1}{3}$ (or equivalently, $\ddot{a} \sim \dot{H} + H^2 < 0$), i.e. when the universe undergoes through a deceleration era; (2) the reversible case gets satisfied for $\omega_\mathrm{eff} = -\frac{1}{3}$, i.e. at the transitions from accelerating to deceleration (or vice-versa); (3) for $\omega_\mathrm{eff} < -\frac{1}{3}$, i.e. when the universe passes through an accelerating stage, neither reversible nor irreversible cases seem to hold --- this may be due to the fact that an accelerating stage of the universe corresponds to a violation of strong energy condition, owing to which, the normal thermodynamic law of the $matter~fields$ needs to be modified. The investigation of the second law of thermodynamics during $acceleration~era$ of the universe is worthwhile to study which we expect to do in some future work.

Therefore we may argue that the reversible (or irreversible) cases of the second law of thermodynamics, in modified gravity (cosmology), gets naturally validated for $\omega_\mathrm{eff} \geq -\frac{1}{3}$ with the proper horizon entropy ($S_\mathrm{h}$) determined in the previous sections.

\section{Conclusion and future remarks}
In conclusion, we reveal the firm inter-connection between gravity and thermodynamics in cosmological context based on the first and second laws of thermodynamics of the apparent horizon. It turns out that the modified gravity theories, from the perspective of the thermodynamic correspondence, can be divided in two categories depending on their independent thermodynamic variables. $Category~I$ (having one thermodynamic d.o.f.): For the class of modified gravity theories where the second FLRW equation contains up to single order time derivative of the Hubble parameter, the independent thermodynamic variable is given by $\left\{H\right\}$; and consequently, the horizon entropy naturally depends only on the Bekenstein-Hawking entropy (for example, the $F(Q)$ or $F(T)$ theory of gravity). Moreover the thermodynamic law, corresponding to such modified theories, proves to be of similar form as in Einstein gravity. Such similarity is due to the fact that both the Einstein gravity and this class of modified gravity theories share the same thermodynamic d.o.f., namely $\left\{H\right\}$, in their equivalent thermodynamic description. On other hand, $Category~II$ (having more than one thermodynamic d.o.f.): for the class of modified gravity theories with higher derivatives of the Hubble parameter in the second FLRW equation (for instance, up to $n$-th order derivative of $H$, with $n>1$), the set of independent thermodynamic variable turns out to be: $\left\{H, \dot{H}, \ddot{H},...., H^{(n-1)}\right\}$; and as a result, the other thermodynamic quantities (including the horizon entropy) depend on these independent variables. Therefore the horizon entropy corresponding to this class of modified theory naturally acquires the time derivative terms of the Bekenstein-Hawking entropy, in particular,
\begin{eqnarray}
 S_\mathrm{h} = S_\mathrm{BH} + \mathrm{terms~containing~up~to~(n-1)~th~order~time~derivative~of~S_\mathrm{BH}} \, .
 \nonumber
\end{eqnarray}
The coefficients of such time derivative terms are proportional to the modification parameter of the gravity theory (compared to the Einstein gravity), which identically vanish and $S_\mathrm{h} \rightarrow S_\mathrm{BH}$ in the limit of Einstein gravity. Moreover the thermodynamic law for the present case of modified theories (with $n>1$) turns out to be of different form than the Einstein gravity (see Eq.~\ref{M-6})). Actually in order to have a consistent thermodynamic description, the Category-II modified theory (with $n>1$) require more thermodynamic d.o.f. compared to the Einstein gravity; and such difference between the Category-II and the Einstein theory gets reflected through the form of their respective thermodynamic law.

Finally we would like to mention some interesting possibilities and some future remarks regarding the correspondence between modified gravity theories with entropic cosmology: (1) By the correspondence between modified gravity theories (cosmology) and entropic cosmology, the current work shows a way how to associate a gravitational Lagrangian (in the gravity sector) corresponding to an entropy of the apparent horizon (in the thermodynamic sector). The associated gravitational Lagrangian in turn leads to Wald entropy which may has important significance in the context of cosmology. Moreover, (3) owing to have an associated Lagrangian with entropic cosmology, we may infer the cosmological perturbation which in turn depicts the GWs as well as the PBH formation in the sector of horizon thermodynamics. These are worthwhile to investigate in future.

 %%%%%%%%%%%%%%%%%%%%%%%%%%%%%%%%%%%%%%%%%%%%%%%%%%%%%%%%%%%%%%%%%%%%%%
 \bibliography{bibliography}
\bibliographystyle{./utphys1}
%%%%%%%%%%%%%%%%%%%%%%%%%%%%%%%%%%%%%%%%%%%%%%%%%%%%%%%%%%%%%%%%%%%%

\section*{Appendix: $F(R,\mathcal{G}) = \frac{R}{16\pi G} + \beta \mathcal{G}^2$ theory and entropic cosmology}\label{sec-appendix}
The gravitational theory like \cite{Nojiri:2010wj,Capozziello:2011et},
\begin{eqnarray}
 \mathcal{A} = \int d^4x\sqrt{-g}\left[\frac{R}{16\pi G} + f(\mathcal{G}) + \mathcal{L}_\mathrm{mat}\right]\, ,
 \label{G-0}
\end{eqnarray}
with $\mathcal{G} = R^2-4R_\mathrm{ab}R^\mathrm{ab}+R_\mathrm{abcd}R^\mathrm{abcd}$ being the Gauss-Bonnet (GB) curvature term (where $a$, $b$, $c$ and $d$ are the four dimensional spacetime index), shows the following FLRW equation:
\begin{eqnarray}
 H^2 + \frac{8\pi G}{3}\left[f(\mathcal{G}) - \mathcal{G}f'(\mathcal{G}) + 24H^3\dot{\mathcal{G}}f''(\mathcal{G})\right] = \frac{8\pi G}{3}\rho \, ,
 \label{G-1}
\end{eqnarray}
for a spatially flat, homogeneous and isotropic metric. Here $(\rho, p)$ represent the energy density and the pressure of the fluid inside the horizon, that satisfies the conservation equation given in Eq.~(\ref{conservation}). It is well known that Einstein-Gauss-Bonnet gravity in four dimensions reduces to standard Einstein gravity --- this may be noted from Eq.~(\ref{G-1}) as $f(\mathcal{G}) = \mathcal{G}$ makes the GB contribution trivial. However for $f(\mathcal{G}) \neq \mathcal{G}$ assists the contribution from the GB term survive. For instance we consider $f(\mathcal{G}) = \beta\mathcal{G}^2$ (with $\beta$ is a constant of mass dimension [-4]), for which, Eq.~(\ref{G-1}) takes the following form,
\begin{eqnarray}
 H^2 + 24^2\beta H^4\left(\frac{8\pi G}{3}\right)\left[3\dot{H}^2 + 6H^2\dot{H} + 2H\ddot{H} - H^4\right] = \frac{8\pi G}{3}\rho\, ,
 \label{G-3}
\end{eqnarray}
where $\mathcal{G} = 24H^2(H^2 + \dot{H})$ for the spatially flat FLRW spacetime. Here, Eq.~(\ref{G-3}) along with the conservation relation of the fluid are treated as the independent cosmological field equations for $F(R,\mathcal{G}) = \frac{R}{16\pi G} + \beta \mathcal{G}^2$ gravity theory, as the other FLRW equations can be determined from these two.

The dependency of $\rho = \rho(H, \dot{H}, \ddot{H})$ in the FLRW eq.~(\ref{G-3}) implies that $\left\{H, \dot{H}, \ddot{H}\right\}$ act as the independent thermodynamic variables in the equivalent thermodynamic description of $F(R,\mathcal{G}) = \frac{R}{16\pi G} + \beta \mathcal{G}^2$ gravity theory. Moreover to examine the possible link between $F(R,\mathcal{G}) = \frac{R}{16\pi G} + \beta \mathcal{G}^2$ and entropic cosmology, let us define a quantity $S$ by the way of Eq.~(\ref{F-5}), and then check whether it can stand as the horizon entropy. By using the FLRW Eq.~(\ref{G-3}), one gets,
\begin{eqnarray}
 dS&=&-\left(\frac{\pi}{G}\right)\left[\frac{2}{H^3} + 24^2\beta\left(\frac{8\pi G}{3}\right)\left\{\frac{12\dot{H}^2}{H} + 36H\dot{H} + 10\ddot{H} - 8H^3\right\}\right]dH\nonumber\\
 &+&24^2\beta\left(\frac{8\pi G}{3}\right)\left[6\dot{H} + 6H^2\right]d\dot{H} + 24^2\beta\left(\frac{16\pi G}{3}\right)Hd\ddot{H}\, .
 \label{G-6}
\end{eqnarray}
Eq.~(\ref{G-6}) suggests that, similar to the $F(R)=R+\alpha R^2$ case, the entity $S$ is not total differentiable and hence it can not be the correct thermodynamic entropy for $F(R,\mathcal{G}) = \frac{R}{16\pi G} + \beta \mathcal{G}^2$ cosmology. A closer observation of Eq.~(\ref{G-6}) shows that $\frac{1}{S_\mathrm{BH}^2}$ acts as an integrating factor in this regard, and thus the quantity which is total differentiable is given by:
\begin{eqnarray}
 -\left(\frac{\pi}{G}\right)\frac{dS}{S_\mathrm{BH}^2}&=&\left[2H + 24^2\beta H^4\left(\frac{8\pi G}{3}\right)\left\{\frac{12\dot{H}^2}{H} + 36H\dot{H} + 10\ddot{H} - 8H^3\right\}\right]dH\nonumber\\
 &+&24^2\beta H^4\left(\frac{8\pi G}{3}\right)\left[6\dot{H} + 6H^2\right]d\dot{H} + 24^2\beta H^4\left(\frac{16\pi G}{3}\right)Hd\ddot{H} \, .
 \label{G-6a}
\end{eqnarray}
Therefore we may write,
\begin{eqnarray}
 \frac{dS}{S_\mathrm{BH}^2} = -d\left(\frac{1}{S_\mathrm{h}}\right) \, ,
 \label{G-6b}
\end{eqnarray}
where $S_\mathrm{h}$ is total differentiable, and is given by:
\begin{eqnarray}
 -\left(\frac{\pi}{G}\right)\frac{dS_\mathrm{h}}{S_\mathrm{h}^2}&=&\left[2H + 24^2\beta H^4\left(\frac{8\pi G}{3}\right)\left\{\frac{12\dot{H}^2}{H} + 36H\dot{H} + 10\ddot{H} - 8H^3\right\}\right]dH\nonumber\\
 &+&24^2\beta H^4\left(\frac{8\pi G}{3}\right)\left[6\dot{H} + 6H^2\right]d\dot{H} + 24^2\beta H^5\left(\frac{16\pi G}{3}\right)d\ddot{H}\, , \, .
 \label{G-6c}
\end{eqnarray}
Eq.~(\ref{F-5}), in terms of the state function $S_\mathrm{h}$, can be expressed by,
\begin{eqnarray}
 TdS_\mathrm{h} = -d\left(\rho V\right) + \left\{\frac{1}{2}(\rho - p) + \frac{1}{2}\left[1 - \left(\frac{3H^2}{8\pi G\rho}\right)^2\right](\rho + p)\left(1 + \frac{2H^2}{\dot{H}}\right)\right\}dV \, ,
 \label{G-6d}
\end{eqnarray}
with $\rho = \rho(H,\dot{H},\ddot{H})$ is shown in Eq.~(\ref{G-1}). Clearly the above expression converges to a form like $TdS_\mathrm{BH} = -d(\rho V) + \frac{1}{2}(\rho - p)dV$ in the limit of Einstein gravity, which is the thermodynamic law used for Einstein theory. Therefore we may regard Eq.~(\ref{G-6d}) to be the correct form of the thermodynamic law of the apparent horizon for $F(R,\mathcal{G}) = \frac{R}{16\pi G} + \beta \mathcal{G}^2$ cosmology, where $T$ is the horizon temperature, $S_\mathrm{h}$ is the state function entropy, and the terms in the rhs represent the internal energy and the work done respectively. Importantly, it may be noted that the form of thermodynamic law is same for both the $F(R)= R+\alpha R^2$ and $F(R,\mathcal{G}) = \frac{R}{16\pi G} + \beta \mathcal{G}^2$ gravity theories (cosmology), in fact, as we showed in Sec.~[\ref{sec-M2}] that such the same form holds true for any general modified gravity theories.

With the thermodynamic law in Eq.~(\ref{G-6d}), the form of horizon entropy ($S_\mathrm{h}$) corresponding to the $F(R,\mathcal{G}) = \frac{R}{16\pi G} + \beta \mathcal{G}^2$ cosmology can be obtained by following the similar procedure as of Sec.~[\ref{sec-FR}]. By doing so, one gets
\begin{eqnarray}
 S_\mathrm{h} = S_\mathrm{BH}\left[1 + \beta_1~\frac{1}{S_\mathrm{BH}^3} + \beta_2~\frac{\dot{S}_\mathrm{BH}}{S_\mathrm{BH}^{7/2}} + \beta_3~\frac{\dot{S}_\mathrm{BH}^2}{S_\mathrm{BH}^4} + \beta_4~\frac{\ddot{S}_\mathrm{BH}}{S_\mathrm{BH}^3}\right]^{-1}  \, ,
 \label{G-6f}
\end{eqnarray}
where we use $S_\mathrm{BH} = \pi/(GH^2)$, and $\beta_i$ are given by,
\begin{eqnarray}
 \beta_1&=&-24^2\beta\left(\frac{8\pi G}{3}\right)\left(\frac{\pi}{G}\right)^3~~~~~~~~;~~~~~~~\beta_2=-24^2\beta\left(4\pi G\right)\left(\frac{\pi}{G}\right)^{5/2}\, ,\nonumber\\
 \beta_3&=&24^2\beta\left(6\pi G\right)\left(\frac{\pi}{G}\right)^2~~~~~~~~~\mathrm{and}~~~~~~~\beta_4=-24^2\beta\left(\frac{8\pi G}{3}\right)\left(\frac{\pi}{G}\right)^2\, .
 \label{G-6g}
\end{eqnarray}
Thus as a whole, Eq.~(\ref{G-6f}) gives the horizon entropy that results to the cosmological field equations of $F(R,\mathcal{G}) = \frac{R}{16\pi G} + \beta \mathcal{G}^2$ from the thermodynamics of the apparent horizon, or equivalently, one may argue that the cosmology of $F(R,\mathcal{G}) = \frac{R}{16\pi G} + \beta \mathcal{G}^2$ has a correspondence with entropic cosmology provided that the horizon entropy is given by Eq.~(\ref{G-6f}). It may be noted that the horizon entropy(ies) for both $F(R) = R + \alpha R^2$ and $F(R,\mathcal{G}) = \frac{R}{16\pi G} + \beta \mathcal{G}^2$ theories depends up to second order time derivative of the Bekenstein-Hawking entropy --- this is due to the fact that the second FLRW equation for both these gravity theories contain up to third order time derivative of the Hubble parameter, owing to which, $\left\{S_\mathrm{BH}, \dot{S}_\mathrm{BH}, \ddot{S}_\mathrm{BH}\right\}$ act as the independent thermodynamic variables in the equivalent thermodynamic description of these theories. Importantly, we would like to mention that the above statement regarding the independent variables holds true for any general $F(R)$ as well as for general $F(R,\mathcal{G}) = \frac{R}{16\pi G} + f(\mathcal{G})$ theories.

\end{document}